\documentclass[aps,prd,nofootinbib]{revtex4}
\usepackage{graphicx}
\usepackage{amsfonts}
\usepackage{amsmath}
\usepackage{amssymb}

\begin{document}

\title{Galactic microlensing by acoustic Schwarzschild black holes}
\author{G.F. Akhtaryanova}
\email{yu.rose@mail.ru}
\affiliation{Zel'dovich International Center for Astrophysics, Bashkir State Pedagogical University, 3A, October Revolution Street, Ufa 450077, RB, Russia}
\author{R.Kh. Karimov}
\email{karimov_ramis_92@mail.ru}
\affiliation{Zel'dovich International Center for Astrophysics, Bashkir State Pedagogical University, 3A, October Revolution Street, Ufa 450077, RB, Russia}
\author{U.K. Khidirov}
\email{umurzokk@mail.ru}
\affiliation{Zel'dovich International Center for Astrophysics, Bashkir State Pedagogical University, 3A, October Revolution Street, Ufa 450077, RB, Russia}
\author{R.N. Izmailov}
\email{izmailov.ramil@gmail.com}
\affiliation{Zel'dovich International Center for Astrophysics, Bashkir State Pedagogical University, 3A, October Revolution Street, Ufa 450077, RB, Russia}

\date{02 April 2026}

\begin{abstract}
This work explores the application of acoustic black holes as a novel class of lenses in Galactic microlensing, potentially representing dark matter halo objects. While sharing key features like an event horizon, their underlying fluid-dynamical description differs from the vacuum solutions of Einstein's equations, suggesting potentially distinct observational signatures. This work investigates these distinctions by calculating the galactic microlensing predictions for acoustic black holes and comparing them to the standard Schwarzschild case. Using the observed parameters of known black hole candidates (Cygnus X-1, A0620-00, GRO J1655-40) as illustrative lenses, we demonstrate that the acoustic black holes tuning parameter $\xi $ significantly alters key observables. Our results show that an increase in $\xi $ leads to a larger Einstein ring radius, a longer event duration, and a higher peak magnification in the microlensing Paczy\'{n}ski light curves. Furthermore, we find that the microlensing event rate is enhanced for acoustic black holes compared to their Schwarzschild counterparts, with the probability of detection growing with $\xi$. These findings establish galactic microlensing as a promising astrophysical channel for constraining analogue gravity metrics, with the primary effects being potentially detectable in the statistical analysis of current and future microlensing survey data.
\end{abstract}

\maketitle


\section{Introduction}
\label{sec1}
The study of black holes stands as a cornerstone of modern theoretical physics. Direct experimental investigation of their properties, however, is hindered by their extreme and elusive nature. A promising alternative path is explored through analogue gravity models, which replicate the physics of curved spacetime in controlled laboratory settings \cite{Barcelo:2011}. A seminal example is the acoustic black hole, proposed by Unruh \cite{Unruh:1981}, where a fluid flow exceeding the local speed of sound creates an "acoustic horizon" that traps sound waves, mimicking a gravitational event horizon. Recently, there have been many theoretical and experimental advances in various aspects of analogue gravity, such as the superradiation \cite{Braidotti:2022, Ge:2012, Torres:2017}, the quasinormal modes (QNM)  \cite{Torres:2020}, the Lyapunov exponent \cite{Wang:2020}, the thermodynamics \cite{Mondal:2025, Zhang:2016} and the Sagnac effect \cite{Roy:2024} etc.

While acoustic black holes capture essential features like Hawking radiation \cite{Visser:1998, Sarkar:2017, Vieira:2016, Eskin:2019, Eskin:2021, Zhang:2011, Guo:2020, Yan:2023, Steinhauer:2016, Kolobov:2021, Kouniatalis:2025},
they are fundamentally different from classical Schwarzschild black holes. The distinction lies in their origin: acoustic black holes are emergent phenomena described by fluid dynamics in a condensed matter system, whereas Schwarzschild black holes are pure gravitational solutions of Einstein's field equations. This difference in the underlying field equations suggests that their observable signatures could be distinct.

A powerful method for probing the nature of compact objects is gravitational lensing \cite{Nandi:2024, Izmailov:2022, Nandi:2017, Schneider:1992, Bronnikov:2019, Tsukamoto:2012, Ishkaeva:2023, Izmailov:2019, Asada:2017, Kuhfittig:2014, Portnov:2015, Bugaev:2015, Izmailov:2020a, Godani:2023, Tsupko:2009, Tsupko:2012, Bisnovatyi:2008, Izmailov:2020b}. A specific and highly practical form of this is galactic microlensing \cite{Paczynski:1986, Abe:2010, Gao:2023, Tsukamoto:2017, Safonova:2002, Lukmanova:2016, Akhtaryanova:2024}. This effect occurs when a compact object (e.g., a star, planet, or black hole) in the Milky Way passes in front of a distant background star. The gravitational field of the lens object magnifies the light from the source, producing a characteristic, transient brightening. The standard profile for such an event is described by the Paczy\'{n}ski light curve, which is smooth, symmetric, and has a specific time-dependent magnification pattern \cite{Paczynski:1996}. Galactic microlensing is a particularly valuable method because it is a directly observable effect, with large-scale surveys like OGLE (Optical Gravitational Lensing Experiment) and MOA (Microlensing Observations in Astrophysics) having detected thousands of such events, allowing for the statistical study of dark, compact objects \cite{Udalski:2015}. In \cite{Sajadian:2020} was shown that the positions of the magnification peaks during microlensing of variable stars can change, and this shift differs in different filters.

This work investigates the microlensing signatures of acoustic black holes and contrasts them with Schwarzschild case with occurs when tuning parameter goes to zero. In recent years, much work has been devoted to gravitational lensing by acoustic black holes and black bounces \cite{Qiao:2023, Molla:2023, Pereira:2025}. For our analysis, we utilized the observational microlensing data of three black hole candidates to provide a concrete contextual framework. Our results demonstrate a key difference: as the tuning parameter $\xi$ of the acoustic black hole increases, the peak brightness of the microlensing event is enhanced. Interestingly, the peak positions of the light curves remain unchanged in the first-order approximations, consistent with the Paczy\'{n}ski profile, although subtle deviations emerge when second-order effects are considered. Furthermore, we have investigated the probabilistic characteristics of microlensing events, namely the optical depth and the event rate. Our analysis shows that for acoustic black holes, both the optical depth and the event rate are predicted to be higher than for their Schwarzschild counterparts of equivalent mass, suggesting a potentially greater detectability within a given population.

The paper is organized as follows. In Sec. 2 we shall briefly outline acoustic black holes. Sec. 3 works out microlensing observables. In Sec. 4, we deal with the probabilistic characteristics of microlensing and conclude in Sec. 5. We choose units such that $G = 1$, $c = 1$.

\section{Acoustic black hole}
\label{sec2}
Acoustic black holes can be proposed in several ways \cite{Unruh:1981, Torres:2020, Visser:1998, Ge:2019, Yu:2019, Vieira:2023, Jacquet:2020, Svancara:2024, Vieira:2025a, Vieira:2025b}, not only from laboratory tabletop experiments in condensed matter systems, but also from mechanisms in high-energy physics, astronomy and cosmology. In this section, we elucidate the spacetime geometry of acoustic black holes in curved backgrounds, focusing on the solution originally derived by Ge et al. \cite{Ge:2019}. Clarifying the nature of this metric is essential, as it forms the foundation for our microlensing analysis. It is crucial to emphasize from the outset that this is not merely a laboratory analogue constructed in flat Minkowski spacetime. Instead, it is an exact solution describing a non-vacuum spacetime where a gravitational black hole coexists with an acoustic black hole in the surrounding fluid.

The acoustic black hole in curved spacetime emerges from two complementary high-energy physics frameworks: the relativistic Gross-Pitaevskii theory \cite{Gross:1961, Pitaevskii:1961} and the Einstein-Yang-Mills theory \cite{Ge:2019}. Both approaches yield a unified description of how a flowing superfluid medium imprints an effective geometry on top of a fixed gravitational background.

Starting from the action for a complex scalar field $\varphi$ in curved spacetime,
\begin{equation}
S=\int d^{4}x\sqrt{-g}\left( \left\vert \partial _{\mu }\varphi \right\vert
^{2}+m^{2}\left\vert \varphi \right\vert ^{2}-\frac{\lambda}{2}\left\vert \varphi
\right\vert ^{4}\right),
\end{equation}%
with $m^{2} \sim (T-T_{c})$ and $\lambda$ a coupling constant, one considers perturbations around a background solution $\varphi =\sqrt{\rho _{0}}e^{i\theta _{0}}$. In the fixed spacetime, one could assume the background solution of the scalar field as ($\rho _{0}, \theta _{0}$), then consider the fluctuations around ($\rho _{0}, \theta _{0}$) as $\rho = \rho_{0} + \rho_{1}$ and $\theta = \theta_{0} + \theta_{1}$. Writing $\varphi =\sqrt{\rho_{0}+\rho _{1}}e^{i(\theta _{0}+\theta _{1})}$ and linearizing, the phase fluctuation $\theta _{1}$ satisfies a relativistic wave equation in an effective metric \cite{Ge:2019, Ge:2010}:
\begin{equation}
\frac{1}{\sqrt{-G}}\partial _{\mu }\left( \sqrt{-G}G^{\mu \nu }\partial
_{\nu }\theta _{1}\right) =0.
\end{equation}

The effective metric $G_{\mu \nu }$ encodes both the background spacetime geometry and the fluid four-velocity $v_{\mu }=\partial _{\mu }\theta _{0}$. This construction reveals that the acoustic metric is a Hadamard product of two distinct metric tensors \cite{Ge:2019}: 
\begin{equation}
ds^{2}=\left( g^{GR}\ast g^{ac}\right) _{\mu \nu }dx^{\mu }dx^{\nu },
\end{equation}%
where $g_{\mu \nu }^{GR}$ is the physical spacetime metric (e.g., Schwarzschild) and $g_{\mu \nu }^{ac}$ is the analogue metric arising from the fluid dynamics.

Specializing to a Schwarzschild background spacetime,
\begin{equation}
ds_{bg}^{2}=-\left( 1-\frac{2M}{r}\right) dt^{2}+\left( 1-\frac{2M}{r}%
\right) ^{-1}dr^{2}+r^{2}\left( d\theta ^{2}+\sin ^{2}\theta d\phi
^{2}\right),
\end{equation}%
and assuming a radial fluid velocity profile motivated by free-fall accretion,
\begin{equation}
v_{r}=\sqrt{\frac{2M\xi }{r}},
\end{equation}
one obtains, after working at the critical temperature $\left(m^{2}=0\right) $ and appropriate rescalings \cite{Ge:2019, Guo:2020, Qiao:2023}, the acoustic Schwarzschild black hole metric:
\begin{equation}
ds_{acoustic}^{2}=-f(r)dt^{2}+\frac{dr^{2}}{f(r)}+r^{2}\left( d\theta
^{2}+\sin ^{2}\theta d\phi ^{2}\right) ,
\end{equation}%
where function $f(r)$ is defined as: 
\begin{equation}
f(r)=\left( 1-\frac{2M}{r}\right) \left[ 1-\frac{2M\xi }{r}\left( 1-\frac{2M}{r}\right) \right] .
\end{equation}%
Here, $M$ is the mass parameter, and $\xi \geq 0$ is a dimensionless tuning parameter characterizing the radial fluid velocity. The condition $\xi \geq 0$ ensures the velocity is real. The overall conformal factor $\sqrt{3}c_{s}^{2}$ (with $c_{s}$ the speed of sound) present in the original derivation has been absorbed into the time coordinate or set to unity by choosing without loss of generality, as it does not affect null geodesic structure \cite{Guo:2020, Vieira:2021}.

The acoustic Schwarzschild black hole metric (6)-(7) is not merely an analogy but an exact solution describing a gravitational black hole embedded in a flowing superfluid medium. It features the coexistence of a vacuum event horizon at $r_{s}=2M$ and non-vacuum acoustic horizons $r_{ac\pm }$ arising from the fluid dynamics. Here, $r_{s}=2M$ is the "optical" event horizon, while $r_{ac-}=M(\xi -\sqrt{\xi ^{2}-4\xi })$ and $r_{ac+}=M(\xi +\sqrt{\xi^{2}-4\xi })$ are the interior and exterior "acoustic" event horizons respectively \cite{Ge:2019, Vieira:2022}. The tuning parameter $\xi$ controls the fluid velocity and horizon structure, interpolating between the standard Schwarzschild black hole ($\xi =0$) and a strongly coupled fluid-gravity system ($\xi >4$). This rich structure, with its layered causal regions and thermal properties, makes the acoustic black hole a theoretically compelling object and a promising candidate for observational searches through gravitational lensing and microlensing \cite{Wang:2020, Akhtaryanova:2024, Qiao:2023}.

\section{Microlensing observables}
\label{sec3}
In the weak-field limit $(r\rightarrow \infty )$ the impact parameter $b$ is approximately equal to the closest approach $r$. Thus, the deflection angle for acoustic black hole \cite{Qiao:2023} can be expressed as 
\begin{equation}
\widehat{a}(r) = \frac{4(1+\xi)M}{r} + \frac{3\pi\left(5 +2\xi +5\xi^{2}\right)}{4} \left(\frac{M}{r}\right)^{2} + O\left( \frac{M}{r}\right)^{3}.
\end{equation}%
Note that then $\xi =0$ the bending angle will be as Schwarzchild black hole.

The source angle $\beta $ between the optical axis (the line joining the black hole lens and the observer) and the line joining the observer and the source star can be written from the elementary lensing geometry \cite{Virbhadra:2000, Virbhadra:2002, Keeton:2005} for light traversing the lens on two opposite sides as (see Fig.1 for lensing geometry): 
\begin{equation}
\beta =\frac{b}{D_{L}}-\frac{D_{LS}}{D_{S}}\left[ \frac{4\left( 1+\xi
\right) M}{r}+\frac{3\pi \left( 5+2\xi +5\xi ^{2}\right) }{4}\left( \frac{M}{%
r}\right) ^{2}\right] ,
\end{equation}%
where the Euclidean distances from the observer to the lens are $D_{L}$ and to the source is $D_{S}$, the distance from the lens to the source is $D_{LS}=D_{S}-D_{L}$, respectively.

\begin{figure}[ht!]
\centering
\includegraphics[width=0.75\textwidth]{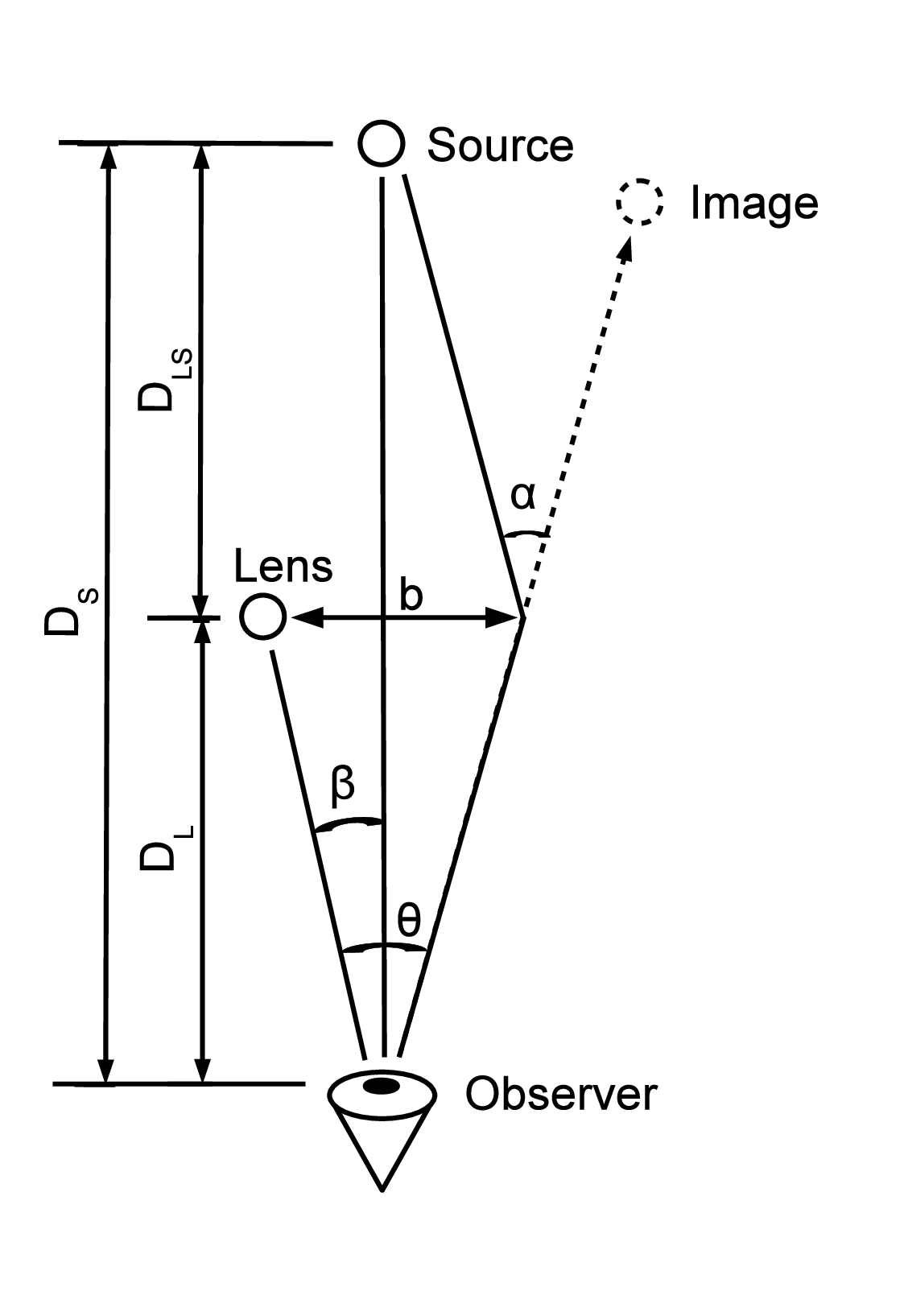}
\caption{Lens geometry, where $D_{L}$ is the distance from the observer to the lens, $D_{S}$ is the distance from the observer to the source, $D_{LS}$ is the distance from the lens to the source, $b$ is the impact parameter of the light, $\protect\beta$ is the angle between lens and source, $\protect\theta$ is the angle between the image and lens, and $\protect\alpha$ is the bending angle between the image and source.}
\label{fig1}
\end{figure}

Since $b\rightarrow r$ as $r\rightarrow \infty$ we obtain 
\begin{equation}
\beta =\frac{r}{D_{L}}-\frac{D_{LS}}{D_{S}}\left[ \frac{4\left( 1+\xi
\right) M}{r}+\frac{3\pi \left( 5+2\xi +5\xi ^{2}\right) }{4}\left( \frac{M}{%
r}\right) ^{2}\right] .
\end{equation}%
If the light source and the lens are exactly aligned with the line of sight (i.e. $\beta =0$), the image is expected to be round (Einstein ring). The Einstein radius $r=R_{E}$, which is defined as the radius of a circular image in the plane of the lens, obtained by solving the cubic equation (10) with $\beta =0$, which gives a positive and negatives roots, but we will take only the positive root 
\begin{equation}
R_{E}=\frac{4}{\sqrt{3}}\sqrt{\frac{M\left( 1+\xi \right) D_{LS}D_{L}}{D_{S}}%
}\sin {\Phi _{1}},
\end{equation}%
where 
\begin{equation}
\Phi _{1}=\frac{\pi }{6}+\frac{1}{3}\arccos {\left[ -\frac{9\sqrt{3}\pi
\left( 5+2\xi +5\xi ^{2}\right) }{64\left( 1+\xi \right) ^{3/2}}\sqrt{\frac{%
MD_{S}}{D_{LS}D_{L}}}\right] }.
\end{equation}

\begin{table}[tbp]
\centering
\begin{tabular}{|c|c|c|c|c|}
\hline
Lens & $r_{s}$ [km] & $\xi $ & $R_{E}$ [km] & $\theta _{E}$ [mas] \\
\hline
Cygnus X-1 \cite{Reid:2011} & $2\cdot 14.8M_{\odot }$ & $0$ & 1.99219$\times 10^{9}$ & 7.01048
\\ 
$D_{S}=8$ kpc & $2\cdot 14.8M_{\odot }$ & $2$ & 3.45057$\times 10^{9}$ & 
12.1425 \\ 
$D_{L}=1.9$ kpc & $2\cdot 14.8M_{\odot }$ & $4$ & 4.45467$\times 10^{9}$ & 
15.6759 \\ 
$D_{LS}=6.1$ kpc & $2\cdot 14.8M_{\odot }$ & $6$ & 5.27084$\times 10^{9}$ & 
18.548 \\
\hline
A0620-00 (V616 Mon) \cite{Eachus:1976} & $2\cdot 11M_{\odot }$ & $0$ & 1.38987$\times 10^{9}$ & 
8.44799 \\ 
$D_{S}=8$ kpc & $2\cdot 11M_{\odot }$ & $2$ & 2.40733$\times 10^{9}$ & 
14.6323 \\ 
$D_{L}=1.1$ kpc & $2\cdot 11M_{\odot }$ & $4$ & 3.10785$\times 10^{9}$ & 
18.8903 \\ 
$D_{LS}=6.9$ kpc & $2\cdot 11M_{\odot }$ & $6$ & 3.67726$\times 10^{9}$ & 
22.3513 \\
\hline
GRO J1655-40 \cite{Orosz:1997} & $2\cdot 7.02M_{\odot }$ & $0$ & 1.57951$\times 10^{9}$ & 
3.30022 \\ 
$D_{S}=8$ kpc & $2\cdot 7.02M_{\odot }$ & $2$ & 2.73579$\times 10^{9}$ & 
5.71615 \\ 
$D_{L}=3.2$ kpc & $2\cdot 7.02M_{\odot }$ & $4$ & 3.53189$\times 10^{9}$ & 
7.37951 \\ 
$D_{LS}=4.8$ kpc & $2\cdot 7.02M_{\odot }$ & $6$ & 4.17899$\times 10^{9}$ & 
8.73156 \\
\hline
\end{tabular}%
\caption{Einstein radius and its angular radius for lens configuration when source taken from the Bulge lensed by black hole located between the observer and the Bulge of the Galaxy.}
\end{table}

Table 1 presents the values $r_{s}$ and $\xi $ of the Einstein radii and angular Einstein radii for a lens configuration where the source is a star in the bulge of a galaxy lensed by a black hole for various objects. Lens parameters, such as its mass and distance, were chosen from catalogs of known compact objects, potential black hole candidates, discovered near our Solar System (Cygnus X-1 \cite{Reid:2011}, A0620-00 (V616 Mon) \cite{Eachus:1976} and GRO J1655-40 \cite{Orosz:1997}). The table shows that for all lens systems, an increase in the parameter $\xi $ leads to an increase in the Einstein radius.

Using the substitution $\frac{r}{D_{L}}=\theta $, equation (10) can be rewritten as 
\begin{equation}
\beta =\theta -4\left( 1+\xi \right) \frac{MD_{LS}}{D_{S}D_{L}}\frac{1}{%
\theta }-\frac{3\pi \left( 5+2\xi +5\xi ^{2}\right) }{4}\frac{M^{2}D_{LS}}{%
D_{S}D_{L}^{2}}\frac{1}{\theta ^{2}}.
\end{equation}%
Using the angular Einstein radius $\theta_{E}=R_{E}/D_{L}$, we obtain 
\begin{equation}
\beta =\theta -4\left( 1+\xi \right) \frac{MD_{LS}}{R_{E}D_{S}}\frac{\theta
_{E}}{\theta }-\frac{3\pi \left( 5+2\xi +5\xi ^{2}\right) }{4}\frac{%
M^{2}D_{LS}}{R_{E}^{2}D_{S}}\frac{\theta _{E}^{2}}{\theta ^{2}}.
\end{equation}%
Eq.(14) can be rewritten as 
\begin{equation}
\frac{\theta ^{3}}{\theta _{E}^{3}}-\frac{\beta }{\theta _{E}}\frac{\theta
^{2}}{\theta _{E}^{2}}-4\left( 1+\xi \right) \frac{MD_{LS}D_{L}}{%
R_{E}^{2}D_{S}}\frac{\theta }{\theta _{E}}-\frac{3\pi \left( 5+2\xi +5\xi
^{2}\right) }{4}\frac{M^{2}D_{LS}D_{L}}{R_{E}^{3}D_{S}}=0.
\end{equation}

Using reduced parameters $\hat{\beta}=\beta /\theta _{E}$ and $\hat{\theta}=\theta /\theta _{E}$, Eq. (15) becomes a qubic equation: 
\begin{equation}
\hat{\theta}^{3}-\hat{\beta}\hat{\theta}^{2}-\mu \hat{\theta}-\nu =0,
\end{equation}%
where $\mu $ and $\nu $ are functions of $\xi $, $M$, $D_{S}$, $D_{L}$, $D_{LS}$ and given as 
\begin{eqnarray}
\mu &=&\frac{4(1+\xi )MD_{LS}D_{L}}{R_{E}^{2}D_{S}}=\frac{3}{4\sin ^{2}{\Phi
_{1}}}, \\
\nu &=&\frac{3\pi \left( 5+2\xi +5\xi ^{2}\right) }{4}\frac{M^{2}D_{LS}D_{L}%
}{R_{E}^{3}D_{S}}=\frac{9\sqrt{3}\pi \left( 5+2\xi +5\xi ^{2}\right) }{%
256\left( 1+\xi \right) ^{3/2}\sin ^{3}{\Phi _{1}}}\sqrt{\frac{MD_{S}}{%
D_{L}D_{LS}}}.
\end{eqnarray}

Solving the cubic Eq.(16) we get two real solutions 
\begin{equation}
\widehat{\theta }_{1}=\frac{\hat{\beta}}{3}+\frac{2}{3}\sqrt{\hat{\beta}^{2}+%
\frac{9}{4\sin ^{2}{\Phi _{1}}}}\sin {\Phi _{2}},\hspace{1cm}(\hat{\theta}%
_{1}>1),
\end{equation}%
\begin{equation}
\widehat{\theta }_{2}=\frac{\hat{\beta}}{3}-\frac{2}{3}\sqrt{\hat{\beta}^{2}+%
\frac{9}{4\sin ^{2}{\Phi _{1}}}}\cos {\left( \Phi _{2}-\frac{\pi }{6}\right) 
},\hspace{1cm}(\hat{\theta}_{2}<1),
\end{equation}%
where 
\begin{eqnarray}
\Phi _{2} &=&\frac{\pi }{6}+\frac{1}{3}\arccos {\left[ \frac{1}{2}\left\{ -2%
\hat{\beta}^{3}-\frac{27\hat{\beta}}{4\sin ^{2}{\Phi _{1}}}+\frac{243\sqrt{3}%
\pi \left( 5+2\xi +5\xi ^{2}\right) }{256\left( 1+\xi \right) ^{3/2}\sin ^{3}%
{\Phi _{1}}}\sqrt{\frac{MD_{S}}{D_{L}D_{LS}}}\right\} \right. }  \notag \\
&&\times \left. \left( \hat{\beta}^{2}+\frac{9}{4\sin ^{2}{\Phi _{1}}}%
\right) ^{-3/2}\right] .
\end{eqnarray}

We use the $A$ brightness increase formula for the light curves, the formula looks like: 
\begin{equation}
A=A_{1}+A_{2}=\left\vert \frac{\widehat{\theta }_{1}}{\widehat{\beta }}\frac{%
d\widehat{\theta }_{1}}{d\widehat{\beta }}\right\vert +\left\vert \frac{%
\widehat{\theta }_{2}}{\widehat{\beta }}\frac{d\widehat{\theta }_{2}}{d%
\widehat{\beta }}\right\vert ,
\end{equation}%
where $A_{1}$, $A_{2}$ are the magnification of the external and internal images. Since the analytical expressions obtained in Eq.(22) are very large, we will present the results graphically.

The lens movement can be described by time dependence as follows \cite{Abe:2010} 
\begin{equation}
\widehat{\beta }=\sqrt{\widehat{\beta }_{0}^{2}+\frac{\left( t-t_{0}\right)
^{2}}{t_{E}^{2}}},
\end{equation}%
where $\widehat{\beta }_{0}$ is the parameter of the impact of the source trajectory, $t_{0}$ is the time of closest approach, $t_{E}$ is the Einstein radius crossing time, given by the formula 
\begin{equation}
t_{E}=\frac{R_{E}}{\vartheta _{T}},
\end{equation}%
where $\vartheta _{T}$ is the transverse velocity of the lens relative to the source and the observer.

\begin{table}[tbp]
\centering
\begin{tabular}{|c|c|c|c|c|}
\hline
Lens & $r_{s}$ [km] & $\xi $ & $t_{E}$ [day] & $t_{E}$ [day] \\ \hline
&  &  & Bound & Unbound \\ \hline
Cygnus X-1 & $2\cdot 14.8M_{\odot }$ & $0$ & 104.808 & 4.61155 \\ 
$D_{S}=8$ kpc & $2\cdot 14.8M_{\odot }$ & $2$ & 181.533 & 7.98744 \\ 
$D_{L}=1.9$ kpc & $2\cdot 14.8M_{\odot }$ & $4$ & 234.358 & 10.3117 \\ 
$D_{LS}=6.1$ kpc & $2\cdot 14.8M_{\odot }$ & $6$ & 277.296 & 12.201 \\ \hline
\multicolumn{1}{|c|}{A0620-00 (V616 Mon)} & $2\cdot 11M_{\odot }$ & $0$ & 
73.1205 & 3.2173 \\ 
\multicolumn{1}{|c|}{$D_{S}=8$ kpc} & $2\cdot 11M_{\odot }$ & $2$ & 126.648
& 5.57253 \\ 
\multicolumn{1}{|c|}{$D_{L}=1.1$ kpc} & $2\cdot 11M_{\odot }$ & $4$ & 163.502
& 7.1941 \\ 
\multicolumn{1}{|c|}{$D_{LS}=6.9$ kpc} & $2\cdot 11M_{\odot }$ & $6$ & 
193.459 & 8.51218 \\ \hline
GRO J1655-40 & $2\cdot 7.02M_{\odot }$ & $0$ & 83.0971 & 3.65627 \\ 
$D_{S}=8$ kpc & $2\cdot 7.02M_{\odot }$ & $2$ & 143.928 & 6.33285 \\ 
$D_{L}=3.2$ kpc & $2\cdot 7.02M_{\odot }$ & $4$ & 185.811 & 8.17568 \\ 
$D_{LS}=4.8$ kpc & $2\cdot 7.02M_{\odot }$ & $6$ & 219.854 & 9.67359 \\ 
\hline
\end{tabular}%
\caption{Einstein radius crossing time for lens configuration when source taken from the Bulge lensed by black hole located between the observer and the Bulge of the Galaxy, where $v_{T}=220$ km/s for bound system and $v_{T}=5000$ km/s for unbound system.}
\end{table}

The possibility of detecting a star's brightness increase depends on the time scale. The Einstein radius crossing time depends on the transverse velocity. Here, we assume that the black hole's velocity is approximately equal to the star's rotational velocity ($\vartheta _{T}=220$ km/s) if it is bound to the Galaxy. If the black hole were not bound to our Galaxy, the transverse velocity would be much higher. We assume $\vartheta _{T}=5000$ km/s \cite{Safonova:2002} for an unbound black hole. Table 2 shows the Einstein radius crossing time for bulge stars. The tables show that as $\xi$ increases, the Einstein ring crossing time increases. For a gravitationally unbound system, this time is significantly shorter.

\begin{figure}[ht!]
\centering
\includegraphics[width=0.9\textwidth]{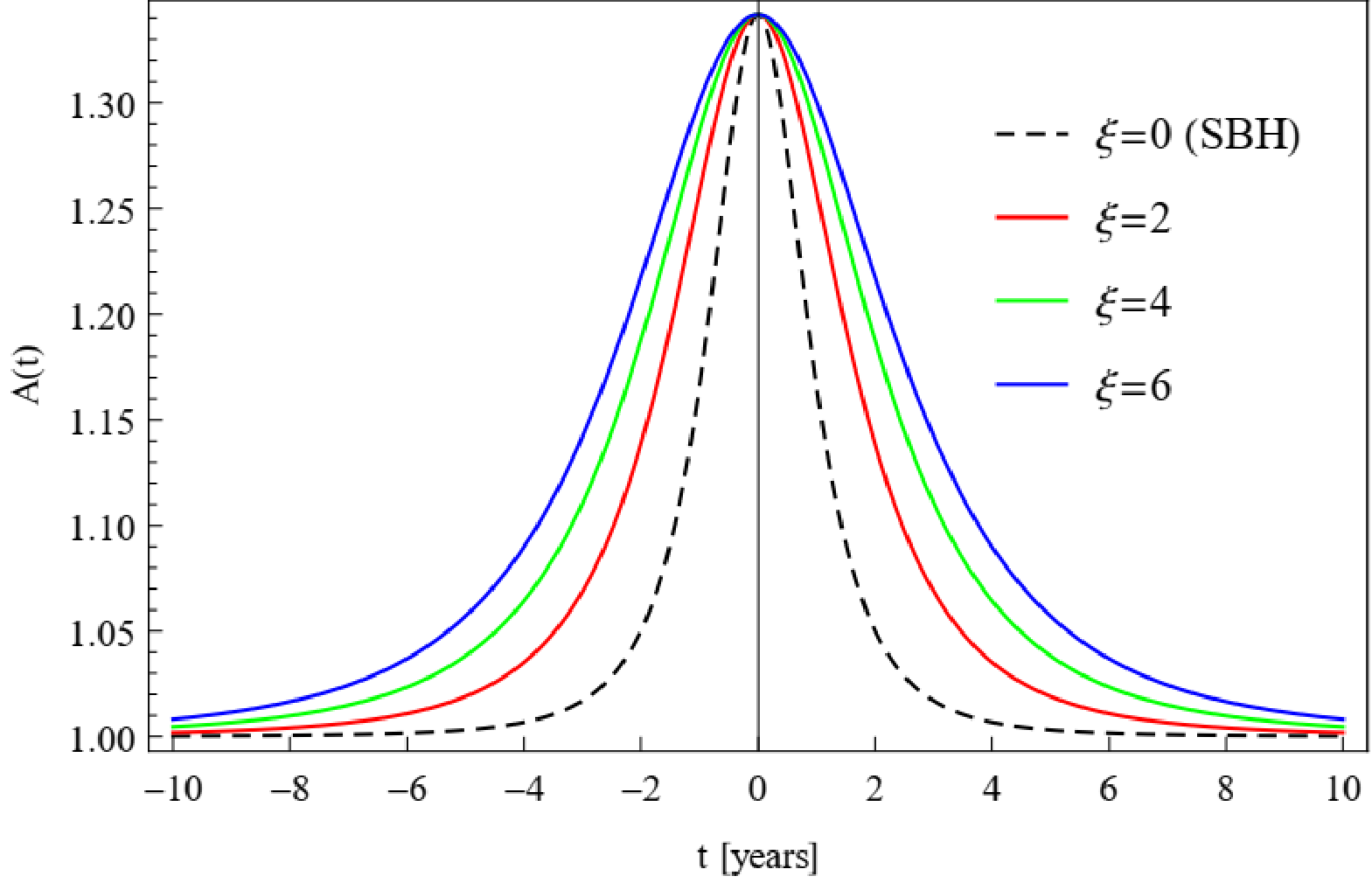}
\caption{Acoustic black hole light curves for $\widehat{\protect\beta }_{0}=1$. The various
values of the tunning parameter $\protect\xi $, indicated as follows: $%
\protect\xi =0$ - dashed black line; $\protect\xi =2$ - solid red line; $%
\protect\xi =4$ - solid green line; $\protect\xi =6$ - solid blue line. It
can be seen from the graph that when we increase the tunning parameter $%
\protect\xi$, the brightness curves of the acoustic black hole increase, but the peaks
remain at one point.}
\label{fig1}
\end{figure}

Paczy\'{n}ski's light curves for acoustic black holes at $\xi =0$, $2$, $4$, and $6$ are shown in Figure 2. As $\xi $ increases, the image brightness increases. Figure 2 shows that the peaks of the curves are independent of the parameter $\xi$. However, if we consider their differences, we find that the peaks of the light curves have a slight deviation. Differences between peaks of light curves: 
\begin{equation*}
A(t)_{\text{max}}|_{\xi =0}-A(t)_{\text{max}}|_{\xi =2}=3.70\cdot 10^{-10},
\end{equation*}%
\begin{equation*}
A(t)_{\text{max}}|_{\xi =0}-A(t)_{\text{max}}|_{\xi =6}=3.59\cdot 10^{-9}.
\end{equation*}

When we look at the second order light curves, minor differences of the order of $10^{-10}-10^{-9}$ appear.

The detection of microlensing events at the level of $10^{-9}$ in flux variations, as predicted by our second-order analysis, presents a significant observational challenge. However, a careful examination of the current state-of-the-art in microlensing surveys and their statistical methodologies reveals that such small deviations, while not directly detectable in individual events, can become accessible through population studies and high-precision photometry. Modern microlensing surveys routinely achieve photometric precision at the level of 1\% or better for bright sources. The OGLE project, which has detected thousands of microlensing events, typically reports photometric errors of 1-2\% for well-sampled light curves \cite{Udalski:2005}. However, detecting deviations of $10^{-9}$ in flux (equivalent to approximately 1 nano-magnitude) is far beyond the capabilities of current ground-based photometry. Even space-based observatories like Kepler and TESS, which achieve parts-per-million precision for bright stars, cannot reach this sensitivity \cite{Sajadian:2024}. The key insight is that such tiny deviations need not be detected in individual events to be scientifically valuable. Modern microlensing science increasingly relies on statistical analysis of large event samples rather than detailed modeling of individual light curves. The microlensing optical depth toward the Galactic bulge, typically of order $10^{-6}$, is determined not from individual events but from the statistical occurrence rate across millions of monitored stars. Mr\'{o}z et al. \cite{Mruz:2019} used 8 years of OGLE-IV observations to measure the optical depth with unprecedented precision, demonstrating how ensemble statistics can extract robust signals from noisy individual measurements.

Therefore, our primary claim is not that individual acoustic black hole events will show $10^{-9}$ deviations, but rather that the population-level statistics of microlensing events -- rates, timescales, and peak magnification distributions -- can reveal the presence of acoustic black hole environments through comparison with the standard Schwarzschild case.

\section{Probabilistic characteristics of microlensing}
\label{sec4}
The probability of a microlensing event to occur for a star is expressed by
the optical depth $\tau$ \cite{Abe:2010}: 
\begin{equation}
\tau =\pi \int\limits_{D_{S}}^{D_{Bulge}}n(D_{L})R_{E}^{2}dD_{s},
\end{equation}%
where $n(D_{L})$ is the number density of black hole as a function of the
line of sight, $D_{Bulge}$ is taken as $8$ kpc.

The event rate expected for a source star $\Gamma$ is defined
by 
\begin{equation}
\Gamma = 2\int\limits_{D_{S}}^{D_{Bulge}}n(D_{L})R_{E}v_{T}dD_{S}.
\end{equation}

For bound system $n=0.147$ pc$^{-3}$ and unbound system $n=4.97\times
10^{-9} $ pc$^{-3}.$

It is assumed that the effective number of stars required to detect acoustic black holes is $10^{6}$. For one or more microlensing events by acoustic black holes to be detected, $\Gamma > 10^{-6}$ is required. Table 3 presents the $\Gamma$ results for different lens types. The lowest probability of such events corresponds to the GRO J1655-40 system, which is about 21 events per year at $\xi = 0$ and 57 events per year at $\xi = 6$. The highest probability of detection corresponds to the Cygnus X-1 system, the probability of detecting such events is from 37 to 100 for $\xi$ from 0 to 6.

Here $\xi$ should be interpreted as an effective environmental parameter: $0 \leq \xi < 4$ corresponds to the more Schwarzschild-like regime, while $\xi > 4$ introduces additional acoustic horizons without implying a pathology; in the Galactic weak-field limit, the leading scalings are $\tau \propto (1 + \xi)$ and $\Gamma \propto \sqrt{1+\xi}$, so order-unity values of $\xi$ already produce a measurable statistical enhancement, whereas $\xi \sim 4 - 6$ gives the strongest effect explored here.

\begin{table}[tbp]
\centering
\begin{tabular}{|c|c|c|c|c|c|c|}
\hline
Lens & $r_{s}$ [km] & $\xi $ & \multicolumn{2}{|c|}{$\tau $} & 
\multicolumn{2}{|c|}{$\Gamma$ [1/year]} \\ \hline
&  &  & Bound & Unbound & Bound & Unbound \\ \hline
Cygnus X-1 & $2\cdot 14.8M_{\odot }$ & $0$ & 2.30824$\times 10^{-5}$ & 
7.80406$\times 10^{-13}$ & 3.72267$\times 10^{-5}$ & 2.86049$\times 10^{-11}$
\\ 
$D_{S}=8$ kpc & $2\cdot 14.8M_{\odot }$ & $2$ & 6.92473$\times 10^{-5}$ & 
2.34122$\times 10^{-12}$ & 6.44785$\times 10^{-5}$ & 4.95452$\times 10^{-11}$
\\ 
$D_{L}=1.9$ kpc & $2\cdot 14.8M_{\odot }$ & $4$ & 1.15412$\times 10^{-4}$ & 
3.90203$\times 10^{-12}$ & 8.32414$\times 10^{-5}$ & 6.39625$\times 10^{-11}$
\\ 
$D_{LS}=6.1$ kpc & $2\cdot 14.8M_{\odot }$ & $6$ & 1.61577$\times 10^{-4}$ & 
5.46285$\times 10^{-12}$ & 9.84925$\times 10^{-5}$ & 7.56815$\times 10^{-11}$
\\ \hline
A0620-00 (V616 Mon) & $2\cdot 11M_{\odot }$ & $0$ & 1.53334$\times 10^{-5}$
& 5.18416$\times 10^{-13}$ & 3.15407$\times 10^{-5}$ & 2.42358$\times
10^{-11}$ \\ 
$D_{S}=8$ kpc & $2\cdot 11M_{\odot }$ & $2$ & 4.60003$\times 10^{-5}$ & 
1.55525$\times 10^{-12}$ & 5.46302$\times 10^{-5}$ & 4.19777$\times 10^{-11}$
\\ 
$D_{L}=1.1$ kpc & $2\cdot 11M_{\odot }$ & $4$ & 7.66671$\times 10^{-5}$ & 
2.59208$\times 10^{-12}$ & 7.05272$\times 10^{-5}$ & 5.4193$\times 10^{-11}$
\\ 
$D_{LS}=6.9$ kpc & $2\cdot 11M_{\odot }$ & $6$ & 1.07334$\times 10^{-4}$ & 
3.62891$\times 10^{-12}$ & 8.34489$\times 10^{-5}$ & 6.4122$\times 10^{-11}$
\\ \hline
GRO J1655-40 & $2\cdot 7.02M_{\odot }$ & $0$ & 9.46125$\times 10^{-6}$ & 
3.1988$\times 10^{-13}$ & 2.16337$\times 10^{-5}$ & 1.66233$\times 10^{-11}$
\\ 
$D_{S}=8$ kpc & $2\cdot 7.02M_{\odot}$ & $2$ & 2.83838$\times 10^{-5}$ & 
9.59641$\times 10^{-13}$ & 3.74706$\times 10^{-5}$ & 2.87923$\times 10^{-11}$
\\ 
$D_{L}=3.2$ kpc & $2\cdot 7.02M_{\odot }$ & $4$ & 4.73063$\times 10^{-5}$ & 
1.5994$\times 10^{-12}$ & 4.83743$\times 10^{-5}$ & 3.71707$\times 10^{-11}$
\\ 
$D_{LS}=4.8$ kpc & $2\cdot 7.02M_{\odot }$ & $6$ & 6.62288$\times 10^{-5}$ & 
2.23916$\times 10^{-12}$ & 5.72373$\times 10^{-5}$ & 4.3981$\times 10^{-11}$
\\ \hline
\end{tabular}%
\caption{The optical depth $\protect\tau$ and event rate $\Gamma$ of a microlensing event by the acoustic black hole when the source star is in the Bulge.}
\end{table}

The extrapolation of the tuning parameter $\xi$ for different astrophysical environments is a critical step in translating the mathematical model of an acoustic black hole into a predictive tool for microlensing observations.

In these models, $\xi$ is a dimensionless parameter that represents the ratio of the fluid's background velocity $v$ to the local speed of sound $c_s$ (i.e., $\xi \approx v/c_s$). By adjusting $\xi$, one can model how different types of "fluid" media (from dense interstellar gas to the plasma in an accretion disk) affect the bending of acoustic or light waves.

\begin{table}[tbp]
\centering
\begin{tabular}{|l|l|c|c|}
\hline
Environment & Fluid Condition & Illustrative effective & Expected Event Rate \\
 & & $\xi$ bins & Enhancement \\
\hline
Accretion Disks & Relativistic/ & $4.0 - 6.0$ & Highest (Statistical excess) \\
 & High-speed flow & & Highest (Statistical excess) \\
Stellar Winds   & Moderate flow & $1.0 - 2.5$ & Moderate (Observable signature) \\ 
Diffuse ISM     & Static/Low-speed & $0.0 - 0.5$ & Lowest (Standard GR behavior) \\
 & flow & & \\ 
\hline
\end{tabular}%
\caption{Relationship of the dimensionless parameter $\xi$ with the type of fluid.}
\end{table}

The Table 4 summarizes how the choice of $\xi$ scales with the environment to ensure detection. By tailoring $\xi$ to the known velocity profiles of these environments, we argue that microlensing monitoring campaigns (targeting $10^6$ stars) can be optimized to look specifically for the high-cadence signatures expected in "fast-fluid" environments like Cygnus X-1.

\section{Conclusion}
\label{sec5}
Galactic microlensing offers a unique astronomical probe into the nature of
compact objects. This study leverages this technique to investigate acoustic
black holes, analogue gravity systems that may exhibit observable signatures
distinct from classical Schwarzschild black holes. We demonstrate that the
acoustic black holes tuning parameter $\xi $, which quantifies deviations
from the Schwarzschild metric, produces pronounced effects in microlensing
events. For a range of astrophysical black hole candidates, we show that
increasing $\xi $ systematically enhances the peak magnification, extends
the event duration, and boosts the predicted microlensing event rate.
Crucially, the first-order light curves retain their general Paczy\'{n}ski
shape, masking the underlying spacetime geometry, while second-order
analysis reveals subtle but characteristic deviations. Our findings provide
specific, quantitative predictions that can be tested against data from
large-scale microlensing surveys. By situating our analysis within the
context of real galactic microlensing surveys and using observed black hole
candidates (Cygnus X-1, A0620-00, GRO J1655-40) as lenses, we have
quantified the observable distinctions that could arise if these objects
were described by the analogue gravity metric.

Qualitative generic effects of tuning parameter $\xi $ on lensing
observables are as follows:

i. The analysis reveals that the fundamental scale of the microlensing
event, the Einstein ring radius $R_{E}$ and its angular size $\theta _{E}$,
increases with the acoustic black hole parameter $\xi $. This directly
translates into a longer Einstein ring crossing time $t_{E}$. The effect is
pronounced, with crossing times being significantly shorter for
high-velocity, gravitationally unbound sources compared to bound ones. This
establishes $\xi $ as a key parameter influencing the geometry and temporal
duration of the lensing event.

ii. The generated Paczy\'{n}ski light curves demonstrate a clear and
measurable signature: the peak magnification increases with the parameter $%
\xi $. While the first-order light curves maintain the characteristic
symmetric shape and peak position of the standard Paczy\'{n}ski profile, a
more detailed analysis reveals subtle deviations at the second order, on the
level of $10^{-10}$ to $10^{-9}$. This indicates that the underlying
spacetime geometry of an acoustic black hole produces a systematically
brighter and subtly different microlensing event compared to a Schwarzschild
black hole of the same mass.

iii.\qquad The calculation of the microlensing event rate $\Gamma $ for a
monitoring campaign of $10^{6}$ stars shows that the detection probability
for acoustic black holes is non-negligible and is enhanced for higher values
of $\xi $. Among the considered candidates, the Cygnus X-1 system yields the
highest predicted event rate, ranging from approximately 37 to 100 events
per year for $\xi $ between 0 and 6. This suggests that, for a given
population, acoustic black holes could be statistically more likely to be
detected via microlensing than their Schwarzschild counterparts.

It is important to contextualize our results within the broader effort to understand black holes immersed in various forms of environmental matter. The metric we study describes a specific, theoretically well-motivated non-vacuum spacetime where a gravitational black hole coexists with an acoustic horizon arising from a flowing superfluid \cite{Ge:2019}. This places our work in parallel with a literature that investigates black holes surrounded by realistic astrophysical environments. Just as the acoustic medium in our model imprints observable signatures on microlensing light curves, these environments are being shown to have an impact on various aspects of black hole physics and observational signatures. For instance, recent studies have developed exact and numerical solutions for black holes embedded in generic matter profiles, including dark matter halos described by Hernquist, Navarro-Frenk-White, and Einasto profiles \cite{Cardoso:2022, Figueiredo:2023, Fonseca:2025}. These works have demonstrated that environmental effects can alter the geodesic structure of spacetime - shifting the innermost stable circular orbit and light ring positions \cite{Fonseca:2025} -- and modify the generation and propagation of gravitational waves \cite{Cardoso:2022, Barausse:2014, Cardoso:2020, Destounis:2023b, Cardoso:2022b}. Furthermore, the presence of matter has been shown to impact dynamical processes like superradiance \cite{Mollicone:2025} and to cause a gravitational
redshift of quasinormal mode QNM frequencies \cite{Pezzella:2025}, an effect analogous to the modifications we find in the microlensing event duration. Crucially, even small environmental perturbations can lead to a spectral instability of black hole QNMs \cite{Cheung:2022, Destounis:2026}, highlighting the sensitivity of strong-field observables to the surrounding medium. Thus the tuning parameter $\xi$ of acoustic black hole may be seen as a specific theoretical realization of a broader phenomenon: the inevitable influence of the environment on observable black hole properties.

In conclusion, our results robustly indicate that acoustic black holes possess unique and identifiable microlensing signatures, characterized by an
enhanced peak magnification, a modified event duration, and a higher predicted event rate. These features provide a concrete, observational test
to distinguish between different black hole metrics. While detecting the subtle second-order effects remains a significant challenge for current
technology, the primary effect of brighter peaks and higher event rates could, in principle, be identified in statistical analyses of large
microlensing datasets. Thus, galactic microlensing emerges as a promising, albeit indirect, astronomical channel to search for and constrain the parameters of analogue gravity systems.

\end{document}